\documentclass[pdftex,twocolumn,epjc3]{svjour3}          % twocolumn

\RequirePackage[T1]{fontenc}

\smartqed  % flush right qed marks, e.g. at end of proof

\RequirePackage{graphicx}
\RequirePackage{mathptmx}      % use Times fonts if available on your TeX system
\RequirePackage{flushend}
\RequirePackage[numbers,sort&compress]{natbib}
\RequirePackage[colorlinks,citecolor=blue,urlcolor=blue,linkcolor=blue]{hyperref}
\def\EDIT#1{{\textcolor{red}{#1}}} 
\usepackage{comment}
\usepackage{soul,csquotes}
\usepackage{amsmath}
\usepackage{booktabs}
\usepackage{subfigure}

\journalname{Eur. Phys. J. C}

\begin{document}

\title{Circular orbits and chaos bound in slow-rotating curved acoustic black holes}

% \subtitle{Do you have a subtitle?\\ If so, write it here}

\author{Balbeer Singh\thanksref{e1,addr1}
        \and
       Nibedita Padhi \thanksref{e2,addr1}
        \and
      Rashmi R. Nayak \thanksref{e3,addr2}%etc.
}

% \thankstext[$\star$]{t1}{Thanks to the title}
\thankstext{e1}{curiosity1729@kgpian.iitkgp.ac.in}
\thankstext{e2}{nibedita.phy@iitkgp.ac.in }
\thankstext{e3}{rashmi@coral.iitkgp.ac.in}

\institute{Department of Physics, Indian Institute of Technology Kharagpur,
 Kharagpur-721302, India\label{addr1}
          %\and
          %Second Address, Street, City, Country\label{addr2}
          \and
          \emph Centre for Ocean, River, Atmosphere and Land Sciences,
Indian Institute of Technology Kharagpur,
Kharagpur-721302, India \label{addr2}}

\date{Received: date / Accepted: date}

\maketitle

\begin{abstract}
% In recent years, the incorporation of acoustic black holes into Schwarzschild spacetime has enabled the simultaneous existence of event and acoustic horizons, as derived from the Gross-Pitaevskii theory. This work investigates the dynamics of a test particle (i.e. vortex) in free fall as it approaches the horizon of a curved slowly-rotating acoustic black hole.
% % , specifically within the context of the  Lens-Thirring spacetime. 
% We analyse the circular orbits of unit mass relativistic test particles around the acoustic black hole
%  and delve into the chaotic motion of the test particle at the unstable equilibrium in the vicinity of the horizon. Furthermore,
%  we explore whether a universal bound exists for the Lyapunov exponent characterising this chaotic motion,
%  drawing parallels to the known bounds for the particle motion in general relativity.
Acoustic black holes, analogs of gravitational black holes created in fluid systems, have recently been embedded within Schwarzschild spacetime using the Gross-Pitaevskii theory, 
% (yielding) 
leading to configurations with both event and acoustic horizons. This study examines the motion of vortices, modeled as unit-mass relativistic test particles, around a slow-rotating curved acoustic black hole. We analyse the stability of circular orbits, identifying the innermost stable circular orbit (ISCO), and investigate the chaotic dynamics of vortices perturbed from unstable circular orbits near the acoustic horizon. Using the Lyapunov exponent to quantify this chaos, we assess whether it satisfies the Maldacena-Shenker-Stanford bound $(\lambda \le 2 \pi T_H)$, a limit established for gravitational black holes in general relativity. Our results show that, in non-extremal cases $(\xi > 4)$, the Lyapunov exponent respects the bound near the horizon, while in extremal cases $(\xi = 4)$, it is violated due to vanishing surface gravity. These findings highlight similarities 
% and differences 
between acoustic and gravitational black holes, advancing the analogy in the context of chaos and orbital dynamics.
\end{abstract}

\section{\label{sec:level1}Introduction}
%\color{blue}
In recent times, analogue black holes have gained significant attention due to their potential to bridge astronomical phenomena with tabletop laboratory experiments. Unruh's groundbreaking work introduced the concept of using hydrodynamic flows as analogous systems to replicate certain aspects of black hole (BH) physics \cite{Unruh:1980cg}.
%The study of analogies has indeed shed some light on the understanding of black holes using the tabletop experiments.
 %and this analogy is fascinating enough to offer a tangible laboratory model 
 %for exploring concepts from curved-space quantum field theory. 
Understanding the propagation of acoustic disturbance in a non-homogeneous fluid flow is difficult, however with certain restrictions, it can be tractable by invoking the language of Lorentzian differential geometry \cite{Visser:1998qn,Visser:1997ux,Visser:2010xv,barcelo2011analogue}. Even if the underlying fluid dynamics is Newtonian and the phenomena take place in flat space plus time, the 
  fluctuations (sound waves) are governed by an effective $(3+1)$-dimensional Lorentzian spacetime geometry. This gives us the basis to draw an analogy between the black holes in Einstein's gravity and those in supersonic flows. By considering supersonic fluid flow, researchers have been able to create an acoustic analogue of a black hole known as a “dumb hole”. This analogy has been extended to demonstrate the presence of phononic Hawking radiation from the acoustic horizon, mirroring the phenomenon predicted by Hawking for real black holes. The Hawking temperature 
$T_{H}$ of phonons emitted from an acoustic black hole is given by the relation $kT_{H}=\frac{\hbar g_{H}}{2\pi c_{s}}$ where $c_{s}$ is the speed of sound and $g_{H}$ represents the surface gravity, which is linked to the rate of change in fluid velocity as it crosses the event horizon, further a steeper velocity gradient results in higher surface gravity and, consequently,
an increased temperature of the phononic radiation \cite{Unruh:1980cg,visser1998hawking}. Thus, acoustic black holes offer an accessible model within fluid dynamics that retains essential black hole properties.\\
\newline
The analog model has recently emerged as a prominent tool in condensed matter physics. Due to significant challenges in directly testing key aspects of general relativity experimentally, the use of condensed matter systems, such as Bose–\\Einstein condensates (BEC), to replicate specific general relativistic phenomena is of considerable importance \cite{Garay:1999sk,Garay:2000jj,Barcelo:2000tg}. The current and most promising analog gravity experiments are performed with fluids and superfluids \cite{Weinfurtner:2010nu,Richartz:2014lda,euve2016observation,Patrick:2018orp,Torres:2020tzs}, Bose-Einstein condensates\cite{Garay:2000jj,Gooding:2020scc,MunozdeNova:2018fxv} and optical media \cite{Drori:2018ivu}. Through these remarkable experiments, the phenomena of Hawking radiation \cite{Weinfurtner:2010nu,Belgiorno:2010wn}, BH superradiance \cite{Richartz:2014lda,Torres:2020tzs}, BH ringdown and QNMs \cite{Patrick:2018orp,Torres:2019sbr}, as well as the cosmological redshift, Hubble friction \cite{Torres:2020tzs} and cosmological pair creation \cite{Wittemer:2019agm} have been observed. In a recent paper \cite{Ge:2015uaa}, the authors demonstrated that acoustic black holes in normal fluids systems can be holographically mapped to real gravitational objects, using black holes in asymptotically Anti-de Sitter (AAdS) spacetime. Furthermore, the geometric configuration of an acoustic black hole within a viscous fluid, accounting for dissipation effects, was analyzed in \cite{Balbinot:2019mei}, and the particle production near the horizon of the acoustic black hole was studied by \cite{Bittencourt:2018yrc}. \\
\newline
The Lyapunov exponent—which quantifies the rate of chaos by measuring the exponential divergence of trajectories \cite{robinson1998dynamical}
—saturates the  Maldacena, Shenker, and Stanford (MSS) chaos bound \cite{maldacenaBoundChaos2016b}, given by ($\hbar=1$) 
\begin{align}\label{bound}
    \lambda \leq 2\pi T_{H},
\end{align}
 where $T_{H}$ is the Hawking temperature, a feature common to horizons at finite temperature. 
% Given that acoustic black holes possess an analogous horizon with a similar geometric character, defined by the transition to supersonic flow, we are motivated to investigate whether this universality holds in analogue systems.
  % The Lyapunov exponent $\lambda$ quantifies the rate of separation of infinitesimally close trajectories in dynamical systems.} Notably, Maldacena, Shenker, and Stanford proposed an upper bound (known as MSS bound) on this exponent for a quantum thermal system \cite{maldacenaBoundChaos2016b}, which is given by
  % \begin{equation}\label{bound}
    % \lambda \le 2 \pi T_{H},
  % \end{equation} 
  % where $T_{H}$ is the Hawking temperature i.e. Lyapunov exponent $\lambda$ is bounded above by and proportional to temperature $T_{H}$. 
  % The bound was originally proved using the shock waves near the horizon of a black hole.
  A plethora of research has been conducted on the Lyapunov exponent and chaos bound by analysing the geodesic motion of the probe particle \cite{PhysRevD.50.R618,PhysRevD.55.4848,Pradhan:2013bli,hashimotoUniversalityChaosParticle2017a,Dalui:2018qqv,Giataganas:2021ghs,gwakViolationBoundChaos2022d,PhysRevD.98.124001,PhysRevD.104.046020,Kan:2021blg,Jeong:2023hom,leiCircularMotionCharged2022,Chen:2022tbb, Xie:2023tjc,Wang:2023crk,Zhao:2018wkl}.
%, where describe how perturbations in the initial conditions of a particle’s trajectory in the near horizon lead to chaotic motion.Giataganas:2024hil
% 
 % The validity of the chaos bound has also been examined in various black hole backgrounds beyond the AdS/CFT correspondence. For instance, as argued in  \cite{ma2022chaotic}, the generalized MSS inequality is upheld in asymptotically flat backgrounds.
 Further, 
 % In addition to the AdS/CFT correspondence, significant insights have emerged on potential connections between gravity and quantum mechanics including 
  Susskind's proposal for GR = QM conjecture \cite{Susskind:2017ney} suggests that exponential growth in operator size corresponds to the increasing radial momentum of a particle as it falls towards a black hole horizon on the gravity side \cite{Susskind:2018tei}. The momentum growth associated with unstable orbits can be quantitatively assessed using the Lyapunov exponent \cite{Ageev:2018msv,Brown:2018kvn}.
 % For unstable orbits, the momentum growth can be quantified by the Lyapunov exponent. Building on this idea, several studies have been conducted \cite{Ageev:2018msv,Brown:2018kvn}.
 Furthermore, in the context of acoustic black holes, it has been shown that for a vortex freely falling toward the horizon, the Lyapunov exponent related to its momentum growth reaches the chaos bound \cite{wangGeometryOutsideAcoustic2020}. This work further explores the universality of bound by examining vortices as relativistic particles around an acoustic black hole embedded in a Schwarzschild-AdS spacetime. 
 Thus, the analogy between gravitational black holes and acoustic black holes extends beyond the phenomenon of Hawking radiation to probe deeper aspects of black hole physics, such as quantum chaos and information scrambling \cite{Susskind:2018tei,Susskind:2020gnl}.\\
 \newline
 % Motivated by these facts, we are interested in considering an analogue acoustic background of a slowly rotating black hole geometry \cite{1918PhyZ...19..156L,bainesPainleveGullstrandFormLenseThirring2020,Vieira:2021ozg}. The importance of the Lense-Thirring metric lies in the fact that handling the full Kerr black hole metric is somewhat difficult \cite{bainesPainleveGullstrandFormLenseThirring2020,Misner:1973prb} and also in case of an actual rotating planet or star, the vacuum solution outside its surface is not precisely Kerr; it is only asymptotically Kerr, so the only region one can consider is the asymptotic region where it reduces to the Lense-Thirring metric which shows the astrophysical relevance of the black hole. Our interest lies in the recently developed Lense-Thirring acoustic black hole (LTABH) derived from the Gross-Pitaevskii and Yang-Mills theory in \cite{vieira2021quasibound}. 
 Motivated by these considerations, in this work, our interest lies in the recently developed slow-rotating acoustic black hole derived from the Gross-Pitaevskii and Yang-Mills theory \cite{1918PhyZ...19..156L,Vieira:2021ozg}. The importance of the Lense-Thirring metric lies in the fact that handling the full Kerr black hole metric is somewhat difficult \cite{bainesPainleveGullstrandFormLenseThirring2020,Misner:1973prb} and also in case of an actual rotating planet or star, the vacuum solution outside its surface is not precisely Kerr; it is only asymptotically Kerr, so the only region one can consider is the asymptotic region where it reduces to the Lense-Thirring metric which shows the astrophysical relevance of the black hole.
To our knowledge, no prior research has applied geodesic circular orbit stability analysis to explore the chaos bound in the context of acoustic black holes. Thus, our primary objective is to derive the Lyapunov exponent and subsequently examine whether the chaos bound holds for the slowly rotating acoustic black hole, employing the approach outlined in \cite{Gao:2022ybw,Wang:2023khb}.\\
\newline
The structure of the paper is as follows. In section \ref{overview} we give a brief overview of 
% the Gross-Pitaevskii theory and review of the derivation of a 
the slow-rotating acoustic black hole.
% from the related Lens-Thirring metric.
In section \ref{Veff and isco} we obtain the effective potential of the particle-like vortex in the vicinity of the acoustic black hole and then further find out the innermost stable circular orbit. Then in section \ref{LE} we first develop the necessary Jacobi matrix to find the Lyapunov exponent and related stability of circular orbits followed by the analysis of the chaos bound using numerical methods.  In the last section, we conclude our results with some comments. 

\color{black} 
\section{Set up: Slow-rotating acoustic black hole} \label{overview}
\begin{figure}[h!]
\begin{subfigure}
    \centering
% \begin{minipage}[]{0.48\linewidth}
  \centering
  \includegraphics[width=\linewidth]{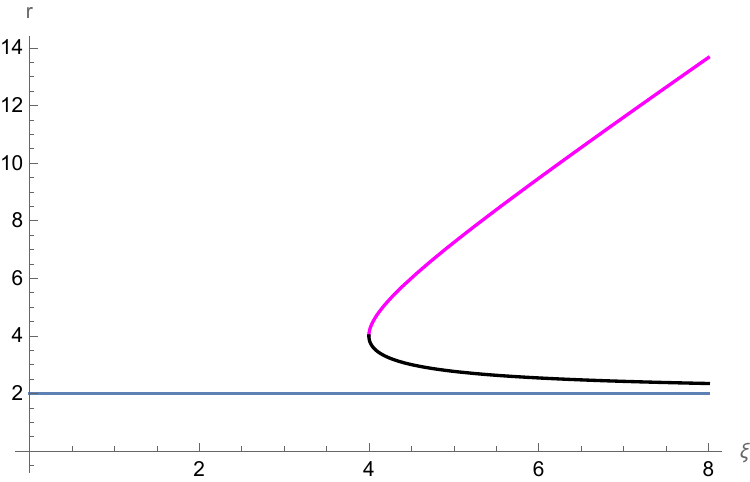}
  % \caption{Plot for Effective potential vs $r$ for $M=1$, $a=0.5$, $\xi=2$, $\epsilon=2$ with $l=0,1,2,3$ (blue, grey, red, magenta).}
  \label{fig:left}
\end{subfigure}
\hspace{0.5cm}
\begin{subfigure}
% \end{minipage}
% \begin{minipage}[]{0.48\linewidth}
  \centering
  \includegraphics[width=\linewidth]{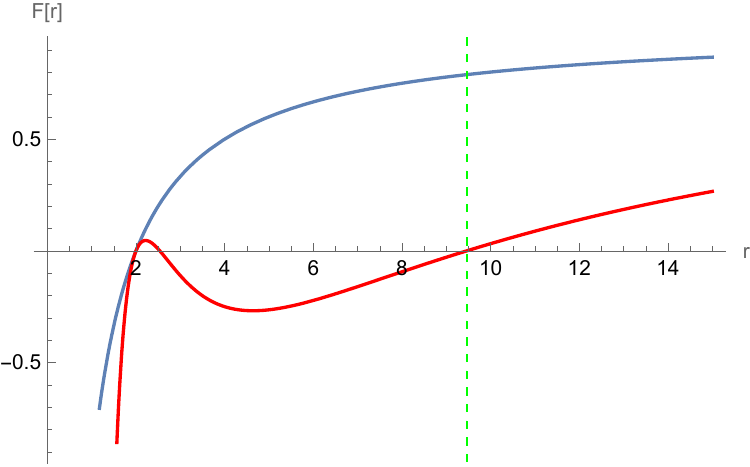}
  %\caption{$V_{\text{eff}}$ at $M=1$, $a=0.5$, $\xi=6$, $\epsilon=2$ with $l=0,1,2,3$ (blue, grey, red, magenta).}
  \label{fig:right}
\end{subfigure}
% \end{minipage}
\caption{(Top panel) Plot for $r$ vs $\xi$ at fixed parameters $M=1$. The blue line represents the $r_{s}$. The (bottom panel) $\mathcal{F}(r)$ vs $r$ at same fixed parameters with $\xi= 0,~ 6$(blue, red). The green dotted curve depicts outermost acoustic horizon, $r_{1}=9.4641.$}
 \label{r-xi-F}
\end{figure}

% Next, we specifically turn our attention to the Lens-Thirring acoustic black hole obtained in 
The metric of the slow-rotating black hole is given by \cite{Vieira:2021ozg}
\begin{align}
     ds^2= - f(r) dt^2 + \frac{1}{f(r)} dr^2 + r^2 d\theta^2 + r^2 \sin^2\theta d\phi^2 \nonumber\\-
    \frac{4a M \sin^2\theta}{r} dt d\phi,
\end{align}
  % \end{eqnarray}
where, $a$ is the rotation parameter  with the limit $a^2 \approx 0$, and the function $f(r)= 1 - \frac{2 M}{r}$. Note that, as mentioned in \cite{Vieira:2021ozg}, after the essential rescaling of the four-velocity $v_{\mu}$ and mass, in the critical temperature limit $(T \xrightarrow{} T_{c})$, we obtain $v^\mu v_{\mu}=-1$.
For the normalisation condition of the fluid four-velocity $v^\mu v_{\mu}=-1$ to follow, the radial velocity and the temporal velocity are considered as 
\begin{eqnarray}
    v_{r}= \sqrt{\frac{2 M \xi}{r}}, ~~v_{t}= \sqrt{f(r) + \frac{2 M \xi}{r} f^{2}(r)},
\end{eqnarray}
where $\xi >0$ is a tuning parameter chosen so that the acoustic event horizon lies outside the event horizon.
%using eq (\ref{big-ABH}).
%by assuming a simple background such that $g_{tt}(r)g_{rr}(r)=-1$ and taking the background phase $\Theta_{0}$ to be independent of the coordinates $\theta, \phi$ which leads to $v_{\theta}=0 = v_{\phi}$. Further, we shall only deal with the critical temperature of the GP theory so that $v_{\mu} v^{\mu}=-2 c_{s}^2$ with the coordinate transformations
%\begin{eqnarray}
 %   dt \xrightarrow{} dt + \frac{v_{r} v_{t}}{g_{tt} \left(c_{s}^{2} - v_{i}v^{i} dr\right)}, \\
%  d\phi \xrightarrow{} d\phi - \frac{g_{t \phi} \left(c_{s}^2 - v_{t}v^{t} \right) v_{t} v_{r} dr}{g_{\phi \phi} \left(c_{s}^2 - v_{\mu}v^{\mu} \right) g_{tt}\left(c_{s}^2 - v_{r}v^{r} \right)},
% \end{eqnarray}
Then, the slow-rotating acoustic black hole metric 
% obtained in eq (\ref{big-ABH}) reduces to
takes the following form \cite{Vieira:2021ozg}
%\begin{eqnarray}
 %   ds^2_{ac}= 
%\end{eqnarray}
%We now fix the spacetime background as the slow-rotating black hole \ref{} 
%\begin{eqnarray}
%\end{eqnarray}
%The metric of the slow-rotating ABH is:
\begin{align}\label{slow-rot}
    ds^2 &= \sqrt{3} c_{s}^2 \Bigg[ - \mathcal{F}(r) dt^2 + \frac{1}{\mathcal{F}(r)} dr^2 + r^2 d\theta^2 + r^2 \sin^2 \theta d \phi^2 \nonumber \\ \quad &
    - \frac{4 M a \sin^2\theta}{r} dt d\phi  \Bigg],
\end{align}
where, 
\begin{equation} \label{eqF}
    \mathcal{F}(r) = f(r) \Bigg[ 1 - f(r)\frac{2 M \xi}{r} \Bigg],
\end{equation}
% Note that as mentioned in \cite{Vieira:2021ozg} after the essential rescaling of the four-velocity in the critical temperature limit, the radial velocity and the temporal velocity are considered as 
% \begin{eqnarray}
%     v_{r}= \sqrt{\frac{2 M \xi}{r}}, ~~v_{t}= \sqrt{f(r) + \frac{2 M \xi}{r} f^{2}(r)},
% \end{eqnarray}
%Without loss of generality, we shall assume $c_{s}= \frac{1}{\sqrt{3}}$ throughout the rest of the paper.
For a spatially two-dimensional model, we choose $\theta= \frac{\pi}{2}$ which defines an equatorial plane.
% \color{red}
Studying motion in the equatorial plane provides valuable insights into the dynamics of the system including the role of the Carter constant
and frame-dragging (see \ref{APPENDIX}). However, it is a restricted scenario and does not fully capture the complexities of the non-spherically symmetric acoustic Kerr metric.
\color{black}
After taking $\theta= \frac{\pi}{2}$, the metric in eq (\ref{slow-rot}) reduces to
\begin{equation} \label{ABH}
      ds^2 = \sqrt{3} c_{s}^2 \Bigg[ - \mathcal{F}(r) dt^2 + \frac{1}{\mathcal{F}(r)} dr^2 + r^2 d\phi^2 - \frac{4 M a}{r} dt d\phi  \Bigg],
\end{equation}
The metric in the matrix form can be rewritten as
\begin{equation}
g_{\mu \nu}=\left(
\begin{array}{ccc}
 -\mathcal{F}(r) & 0 & -\frac{2 a M}{r} \\
 0 & \frac{1}{\mathcal{F}(r)} & 0 \\
 -\frac{2 a M}{r} & 0 & r^2 \\
\end{array}
\right),
\end{equation}
% \begin{equation}
% \mathcal{F}=0=(r-r_{s})(r-r_{1})(r-r_{2}),
% \end{equation}
% \newline
On equating, $\mathcal{F}=0=(r-r_{s})(r-r_{1})(r-r_{2})$, we obtain the outer and inner acoustic horizons denoted by, respectively, $r_{1}= M(\xi + \sqrt{\xi^2 - 4 \xi}), ~~ r_{2}= M(\xi - \sqrt{\xi^2 - 4 \xi})$ whereas the optical event horizon is given by $r_{s}= 2 M$. 
%We can treat $r_{2}$ as the inner acoustic horizon and $r_{1}$ as the outer acoustic horizon,
For the acoustic horizons to exist, we must have $\xi \ge 4$, for the region of parameters falling in $0 \le \xi < 4$ only the optical horizon exists and in this case both the interior and exterior acoustic event horizons vanish. Further when the tuning parameter $\xi=4$, the inner and outer horizons coincide and we obtain the extreme Lens-Thirring acoustic black hole (LTABH)
%which coincides when the tuning parameter $\xi=4$ 
(Fig \ref{r-xi-F}). But, taking the parameter $\xi > 4$, there exist three regions: In the region where $r<r_{s}$,  both light rays (photons) and sound waves (phonons) are unable to escape from the acoustic black hole. For $r_s < r < r_{1}$, light rays can escape from the acoustic black hole, while sound waves cannot. Beyond $r > r_{1}$, both light rays and sound waves have the potential to escape from the acoustic black hole. 
% \cite{guoAcousticBlackHole2020a}.
Moreover, when $\xi=0$, the metric reduces to the original Schwarzchild metric (blue curve in Fig 1). As  $\xi \xrightarrow{} \infty$ the entire spacetime resides within the acoustic black hole. We will henceforth regard the outer acoustic horizon as the acoustic horizon. \\
The surface gravity of the slowly rotating acoustic  black hole and the angular velocity corresponding to the outer horizon respectively, can be expressed as \cite{Vieira:2021ozg}
%for small $a$ 
\begin{eqnarray}
    \kappa &=& \frac{1}{r_{1}^2} \left.\mathcal{F}^\prime(r)\right\vert_{r= r_{1}} = \frac{(r_{1} - r_{2})(r_{1} - r_{s})}{2 r_{1}^2} \label{kappa}, \\
    \Omega&=& \frac{2 M a}{r_{1}^2},
\end{eqnarray}
and therefore the Hawking temperature is given by
\begin{eqnarray}
    T_{H} = \frac{\kappa}{2 \pi k_{B}},
\end{eqnarray}
where $k_{B}$ is the Boltzmann constant.

% \begin{figure*}
% \includegraphics{fig_2}% Here is how to import EPS art
% \caption{\label{fig:wide}Use the figure* environment to get a wide
% figure that spans the page in \texttt{twocolumn} formatting.}
% \end{figure*}
% \begin{minipage}[b]{0.48\linewidth}
%   \centering
%   \includegraphics[width=\linewidth]{Veff-M-1-all-varying.pdf}
%   % \caption{Plot for Effective potential vs $r$ for $M=1$, $a=0.5$, $\xi=2$, $\epsilon=2$ with $l=0,1,2,3$ (blue, grey, red, magenta).}
%   \label{fig:left}
% \end{minipage}
% \hspace{0.5cm}
% \begin{minipage}[b]{0.48\linewidth}

% \begin{figure}[ht!]
% \begin{subfigure}
%   \centering
%   \includegraphics[width=0.45\textwidth]{r-xi.pdf}
% %\put(-210,135){(a) \hspace{2cm} r(0)=11}
%  % \phantomsubcaption\label{fig:nuvr}
% \end{subfigure}
% %\hspace{2.6mm}
% \begin{subfigure}
%   \centering
%   \includegraphics[width=0.45\textwidth]{F-r.pdf}
% %\put(-210,138){(b) \hspace{2cm} r(0)=11}
%   %\phantomsubcaption\label{fig:varnuvsr}
% \end{subfigure}
% \caption{(Left panel) Plot for $r$ vs $\xi$ at fixed parameters $M=1$. The blue line represents the $r_{s}$. The (right panel) $F(r)$ vs $r$ at same fixed parameters with $\xi= 0,~ 6$(blue, red). The green dotted curve depicts outermost acoustic horizon, $r_{1}=9.4641.$}
% \label{r-xi-F}
% \end{figure}

\section{Circular orbits and ISCO} \label{Veff and isco}
 In this section, we turn our attention to the motion of the \enquote{test particle} of unit mass around the slowly rotating acoustic black hole. It was postulated that vortices could act like relativistic particles with their motion dictated by the acoustic metric such that they do not exceed the speed of sound in an almost incompressible fluid \cite{wangGeometryOutsideAcoustic2020,Ge:2019our,guoAcousticBlackHole2020a,Qiao:2021trw,volovik2003universe,Zhang:2004hrz}. The analogue model can be constructed by considering the acoustic propagation in an irrotational vortex \cite{ Zhang:2004hrz}. The dynamics of these particles are influenced by the fluid metric, and their stability is ensured by a topological quantum number when regarded as topological defects. Therefore we shall consider an irrotational vortex
 %particle-like vortex (i.e. irrotational) 
 of unit mass in the presence of the acoustic black hole. We first find the effective potential and then the presence of the stable (and unstable) circular orbits. Towards the end, we also obtain the innermost stable circular orbit (ISCO).
  % In this section, we turn our attention to the motion of the \enquote{test particle}
   %(vortex) 
%   of unit mass around the slowly-rotating acoustic black hole. It was postulated that \enquote{test particles} could act like relativistic particles with their motion dictated by the acoustic metric such that they do not exceed the speed of sound \cite{Zhang:2004hrz,1973JETP...37..341P,PhysRevB.49.12381,wangGeometryOutsideAcoustic2020}. Therefore we shall consider an 
   %particle-like vortex
 %  (irrotational) particle of unit mass in the presence of the acoustic black hole. We first find the effective potential and then the presence of the stable (and unstable) circular orbits. Towards the end, we also obtain the innermost stable circular orbit (ISCO).
%In this section, we shall study the momentum growth of the trajectories of vortices as the test particle near the acoustic horizon of the black hole. We shall take the different initial energy and angular momentum. We will plot and numerically see the growth of the Rindler momentum of the particles as a function of time t.

\subsection{Effective Potential and Circular Orbits}
The metric in eq (\ref{ABH}) is independent of the coodinates $t$ and $\phi$, therefore there are conserved quantities along these directions. If the Killing vectors for our metric are $\xi^{\alpha}= (1,0,0)$ and $\eta^{\alpha}= (0,0,1)$ and $u^{\alpha}$ is the three-velocity of the vortex \cite{Hartle:2003yu,Misner:1973prb}, then, energy per unit mass yields
\begin{eqnarray} \label{energy}
    \epsilon=&& - \xi^{\alpha} g_{\alpha \beta} u^{\beta} \nonumber \\ 
        =&& \sqrt{3} c_{s}^2 \mathcal{F}(r) \frac{dt}{d \tau} + \sqrt{3} c_{s}^2 \left(\frac{2 a M}{r}\right) \frac{d \phi}{d \tau},
\end{eqnarray} 
and the angular momentum per unit mass is given by 
\begin{eqnarray}\label{ang}
    l=&& \eta^{\alpha} g_{\alpha \beta} u^{\beta} \nonumber \\
        =&&  -\sqrt{3} c_{s}^2 \left(\frac{2 a M}{r}\right) \frac{d t}{d \tau} + \sqrt{3} c_{s}^2 r^{2} \frac{d \phi}{d \tau},
\end{eqnarray} 
Using the normalization condition $ g_{\alpha \beta} u^{\alpha} u^{\beta}=-1$, we obtain 
\begin{eqnarray}
    - \mathcal{F}(r) (u^{t})^2 - \frac{4 a M}{r} u^t u^{\phi} + \frac{(u^{r})^2}{\mathcal{F}(r)} + r^2 (u^{\phi})^2 &= -\cfrac{1}{\sqrt{3} c_{s}^2}, \nonumber \\
\end{eqnarray}
or, 
\begin{align}\label{ur}
    (u^{r})^2&= \mathcal{F}(r)\Bigg[\frac{-1}{\sqrt{3} c_{s}^2} + \mathcal{F}(r) (u^{t})^{2} + \frac{4 a M}{r} u^{t} u^{\phi} - r^2 (u^{\phi})^2
    \Bigg], \nonumber \\
\end{align}
On solving the eq (\ref{energy}) and eq (\ref{ang}), we obtain the following relation 
\begin{eqnarray}
    u^{t}= \frac{-2 a l M+r^3 \epsilon }{\sqrt{3} c_{s}^2 r^3 \mathcal{F}(r)},
\end{eqnarray}
and,
\begin{eqnarray}
    u^{\phi}= \frac{2 a M \epsilon + l r \mathcal{F}(r)}{\sqrt{3} c_{s}^2 r^3 \mathcal{F}(r)},
\end{eqnarray}
Then putting these values in the eq (\ref{ur}), we obtain
\begin{align}
     (u^r)^2 &=\cfrac{\epsilon  \left(- 4 a l M+r^3 \epsilon \right)-r \mathcal{F}(r) \left(\sqrt{3} c_{s}^2 r^2+l^2\right)}{3 c_{s}^4 r^3}, \\
          &= \cfrac{\epsilon  \left(-4 a l M+r^3 \epsilon \right)}{r^3}+ \nonumber \\&
           \cfrac{\left(l^2+r^2\right) (2 M -r) \left(r^2-2 M \xi  (r-2 M)\right)}{r^5}, 
\end{align}
%I dont think we need to write eq(29).since eq 28 is in F(r) ,its ok to go with it.
where, in the last equation we have set $c_{s}^2= 1/ \sqrt{3}$. (Note that the pre-factor $\sqrt{3} c_{s}^2$ in the metric acts as the ``conformal factor" \cite{Vieira:2021ozg,Toshmatov:2023fcv}, therefore without loss of generality we have set $c_{s}^2= \frac{1}{\sqrt{3}}$.)
\newline
Thus the effective potential is obtained as 
\begin{align}
    V_{\text{eff}} &= \epsilon^2 - (u^r)^2 \\
            &= \cfrac{1}{r^5} \Big( r^2 \left(4 a l M \epsilon +\left(l^2+r^2\right) (r-2 M)\right) \nonumber \\&
            -2 M \xi  \left(l^2+r^2\right) (r-2 M)^2\Big),
            \label{old-Veff}
\end{align} 
%Therefore the effective potential is  given by 
%\begin{eqnarray}
 %   V_{eff}=&& \frac{\beta + \sqrt{\beta ^2-\alpha  \gamma }}{\alpha } \\
  %       =&&  \frac{2 a l M}{r^3}+ \sqrt{-\frac{\left(l^2+r^2\right) (2 M-r) \left(r^2-2 M \xi  (r-2 M)\right)}{r^5}}
%\end{eqnarray}
which in the limit of $a \xrightarrow{} 0$, and $ \xi \xrightarrow{} 0$ reduces to that of the Schwarzchild \cite{bainesPainleveGullstrandFormLenseThirring2020}.
\\
\begin{figure}[t!]
\begin{subfigure}
\centering
% \begin{minipage}[b]{0.48\linewidth}
  \centering
   \includegraphics[width=\linewidth]{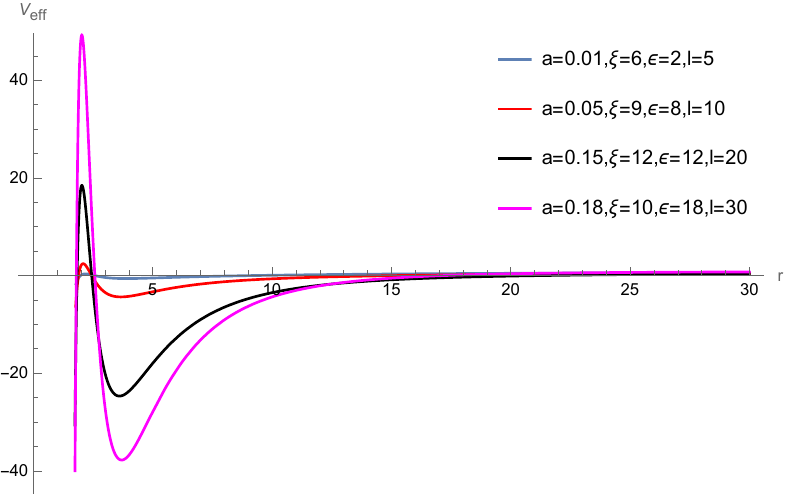}
  %\includegraphics[width=\linewidth]{Veff-M-1-all-varying.pdf}
  % \caption{Plot for Effective potential vs $r$ for $M=1$, and for various parameter values of $a,~ \xi,~ \epsilon$ and $l$.}
   % \caption{Plot for Effective potential vs $r$ for $M=1$, $a=0.01,0.05,0.15,0.18$, $\xi=6,9,12,10$, $\epsilon=2,8,12,18$ with $l=5,10,20,30$ (blue, red, black, magenta).}
  % \label{fig:left}
\end{subfigure}
% \end{minipage}
\hspace{0.5cm}
\begin{subfigure}
% \begin{minipage}[b]{0.48\linewidth}
  \centering
  \includegraphics[width=\linewidth]{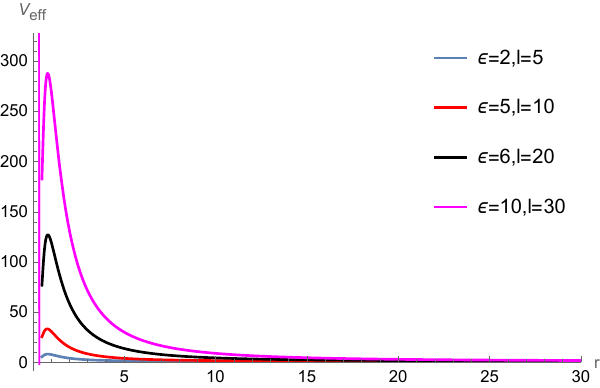}
  %\includegraphics[width=\linewidth]{extremal-Veff-r.pdf}
  % \caption{$V_{\text{eff}}$ at $M=0.1$, $a=0.1$, $\xi=4$, and different values of $\epsilon$ and $l$.}
  % \caption{$V_{\text{eff}}$ at $M=0.1$, $a=0.1$, $\xi=4$, $\epsilon=2,5,6,10$ with $l=5,10.20,30$ (blue, red, black, magenta).}
  % \label{fig:right} 
\end{subfigure}
% \end{minipage}
%\captionsetup{labelformat=empty}
\caption{Comparison of Effective Potential $V_{\text{eff}}$ for different parameters, left panel for non-extremal and right panel for extremal case. (Left panel)Plot for Effective potential vs $r$ for $M=1$, and for various parameter values of $a,~ \xi,~ \epsilon$ and $l$. (Right panel) $V_{\text{eff}}$ at $M=0.1$, $a=0.1$, $\xi=4$, and different values of $\epsilon$ and $l$.}
\label{Veff1}
\end{figure}

\begin{figure}[t!]
\begin{subfigure}
\centering
% \begin{minipage}[b]{0.48\linewidth}
  \centering
    \includegraphics[width=\linewidth]{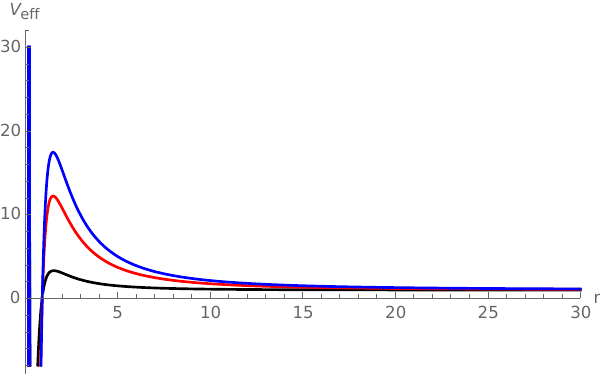}
  %\caption{Plot for Effective potential vs $r$ for $M=1$, $a=0.5$, $\xi=2$, $\epsilon=2$ with $l=0,1,2,3$ (blue, grey, red, magenta).}
  %\label{only-l}
% \end{minipage}
\end{subfigure}
\hspace{0.5cm}
\begin{subfigure}
% \begin{minipage}[b]{0.45\linewidth}
  \centering
  \includegraphics[width=\linewidth]{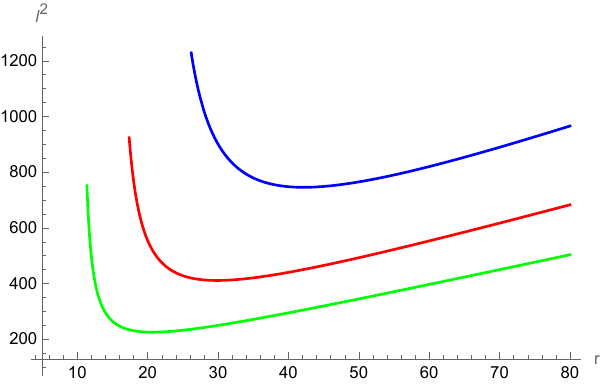}
  %\caption{$V_{\text{eff}}$ at $M=1$, $a=0.5$, $\xi=6$, $\epsilon=2$ with $l=0,1,2,3$ (blue, grey, red, magenta).}
  % \label{}
% \end{minipage}
\end{subfigure}
\caption{(Top panel)Effective potential with $M=0.1, ~\xi=6, ~ a=0.01$ and $\epsilon=10$ at different $l=5,10,12$ (black, red, blue). The (bottom panel) plot showing $l^{2}-r$ plot at parameters $M=1,~ \epsilon=2,$ for pairs $~~a=0.1,0.3,0.08,~~\xi=6,8,4.5$, (blue,red,green).}
 \label{only-l-and-l^2}
\end{figure}

Let us now analyse the circular orbits in the acoustic black hole scenario for which $u^{r}=0$. For a particle to describe the circular orbits we should have
%$$V_{eff}(r_{*})=0$$
%and 
$\left.\frac{d V_{\text{eff}}}{dr}\right\vert_{r= r_{*}}=0 $. Further,
the effective potential for a stable circular orbit should have a local minimum ensured by $ \left.\frac{d^2 V_{\text{eff}}}{d r^2}\right\vert_{r= r_{*}} > 0$, whereas the unstable circular orbit corresponds to the local maximum given by the condition $\left.\frac{d^2 V_{\text{eff}}}{d r^2}\right\vert_{r= r_{*}} < 0$.
 As shown in Fig \ref{Veff1},
 % \ref{fig:left},
%and Figure \ref{Veff2} 
%show the behaviour of the effective potential as a function of r. 
the effective potential attains a minimum value as well as a maximum, therefore there exists stable and unstable circular orbits for the acoustic black hole. However, for the extremal case, we only see the unstable circular orbit (Fig \ref{fig:right}). 
The effective potential acts as a barrier for radial coordinate approaching the near horizon, however, the effective potential quickly becomes zero at the horizon. Further, on analysing for fixed $\xi, ~ M, ~ a$ and $\epsilon$, since the horizon does not depend on $l$, so the effective potential attains the zero at the same position and is independent of the angular number $l$ which is quite evident from the Fig \ref{only-l-and-l^2} (top panel). Moreover, the potential barrier increases on increasing $l$ similar to the Schwarzchild scenario.
\newline
\subsection{Innermost Stable Circular Orbit}
Further on analysing the equation $\cfrac{d V_{\textit{eff}}}{d r}=0$ which gives
\begin{align}
    2 r^2 \left(-6 a l M \epsilon +l^2 (3 M-r)+ M r^2\right) +\nonumber 
    \\ 2 M \xi  (r-2 M) \left(l^2 (3 r-10 M)+r^2 (r-6 M)\right) =0,
\end{align}
and solving for the angular momentum $l^2$, we get\\
% \begin{widetext}
\begin{align}
   % l^{2}&= \frac{1}{\left(-20 M^3 \xi +16 M^2 \xi  r-3 M (\xi +1) r^2+r^3\right)^2} \nonumber \\
   % &\quad \times \left( \left(\sqrt{M r^2 \left(9 a^2 M r^2 \epsilon ^2-\left(20 M^3 \xi -16 M^2 \xi  r + 3 M (\xi +1) r^2-r^3\right) \left(12 M^2 \xi +r (-8 M \xi +\xi  r+r)\right)\right)}+3 a M r^2 \epsilon \right)^2\right),
   l^{2} &= \frac{1}{\left(-20 M^3 \xi +16 M^2 \xi  r-3 M (\xi +1) r^2+r^3\right)^2} \nonumber \\ 
     &\quad \times \Big( \{ M r^2 \Big( 9 a^2 M r^2 \epsilon ^2 \nonumber \\
     &\qquad\qquad - \left(20 M^3 \xi - 16 M^2 \xi  r + 3 M (\xi +1) r^2 - r^3\right) \nonumber \\ % First part before the second times
     &\qquad\qquad \times \left(12 M^2 \xi + r (-8 M \xi +\xi  r+r)\right) \Big) \}^{1/2} \nonumber \\ % Split happens here - line break before this part
     &\qquad\qquad + 3 a M r^2 \epsilon \Big)^{2}, % Optional further break for clarity
\end{align}
% \end{widetext}
 due to the excessively challenging expression of $l$ we obtain the plots of the angular momentum $l^{2}$ with the help of the numerics as shown in Fig \ref{only-l-and-l^2} (bottom panel) for different rotation parameter values.
% \begin{figure*}
%   \centering
%   \includegraphics[width=0.45\textwidth]{l^2-r.pdf}
% %\put(-210,138){(b) \hspace{2cm} r(0)=11}
%  % \phantomsubcaption\label{fig:varnuvsr}
%   \caption{Plot showing $l^{2}-r$ plot at parameters $a=0.1, ~~M=1, ~~\xi=6$, $~ \epsilon=2$.}
% \label{l2plot}
% \end{figure*}
Since the minimum of the angular momentum ensures the presence of the inner stable circular orbit (ISCO) of a test particle, therefore from Fig \ref{only-l-and-l^2} (bottom panel), 
%it is evident that due to the occurrence of the minimum of angular momentum, 
it is evident that there exists an ISCO 
%\cite{Narzilloev:2020qtd} 
\cite{dadhichCircularOrbitsHigher2022,jefremovInnermostStableCircular2015a}. We further plot the $l^2$ vs $r$ plot at different parameters as shown in Fig \ref{only-l-and-l^2} (bottom panel). As $a$ decreases the curve shifts towards the left to smaller $r$. 
\\
The radius of the ISCO in an acoustic rotating black hole depends on the fluid's angular velocity profile which we have 
notice from our analysis and the speed of sound in the fluid. Similar to the Kerr black hole, where the ISCO radius depends on the
spin parameter $a$, in the acoustic case, it will depend on the fluid's rotational properties.
%\color{red} some comment about the interpretation of ISCO in this ABH.
\color{black}
\section{Lyapunov exponent and bound analysis} \label{LE}
% \color{red}
 Unstable circular orbits, when disturbed, exhibit exponential growth in perturbations, which signifies sensitivity to initial conditions \cite{cardosoGeodesicStabilityLyapunov2009a}. 
  % While this instability does not necessarily imply chaos \cite{}, such sensitivity can lead to chaotic behavior in certain cases, particularly in black hole spacetimes \cite{cardosoGeodesicStabilityLyapunov2009a,gwakViolationBoundChaos2022d,hashimotoUniversalityChaosParticle2017a,Dalui:2018qqv}.
 In our work, we have analyzed the Lyapunov exponent as a measure of instability in geodesic motion and used it to explore the potential for chaotic behaviour in the context of black hole dynamics.
Various methods for determining the Lyapunov exponent are discussed in the literature, including \cite{cardosoGeodesicStabilityLyapunov2009a,PhysRevD.98.124001,leiCircularMotionCharged2022,hashimotoUniversalityChaosParticle2017a} as well as approaches like \cite{ gwakViolationBoundChaos2022d,Kan:2021blg}.
\color{black} 
%Cornish:2003ig,In the literature, there exist various techniques to determine the Lyapunov exponent, for instance %\cite{cardosoGeodesicStabilityLyapunov2009a,PhysRevD.98.124001,leiCircularMotionCharged2022,Cornish:2003ig,hashimotoUniversalityChaosParticle2017a} and others such as %\cite{ gwakViolationBoundChaos2022d,Kan:2021blg}. 
%In this section, we will employ the Jacobi matrix method to calculate the Lyapunov exponent \cite{Gao:2022ybw,cornishLyapunovTimescalesBlack2003a}.
Specifically, with the help of the Lagrangian framework, the Lyapunov exponents can be established for particle motion \cite{cardosoGeodesicStabilityLyapunov2009a}. The equations of motion of the particle can be written as 
\begin{align}
    \frac{d y^{i}}{d t} = F_{i}(x^{j}).
\end{align}
On linearising the equations about a certain orbit
\begin{align}
    \frac{d \delta y^{i} (t)}{dt}= K_{ij} \delta y^{j}(t),
\end{align}
where $K_{ij}$ is the Jacobian matrix defined as 
\begin{align}
    K_{ij}= \frac{\partial F_{i}}{\partial y^{j}},
\end{align}
about the solution 
\begin{align}
    \delta y^{i}(t) = L_{ij} \delta y^{i}(0),
\end{align}
where $L_{ij}(t)$ is the evolution matrix satisfying $\dot L_{ij}(t)= K_{il} L_{lj}$ and $L_{ij}(0)= \delta_{ij}$.
 The Lyapunov exponent characterises the average exponential rate of divergence between two nearby trajectories in a dynamical system. It signifies the typical rate of contraction or expansion of nearby orbits in the phase space. The eigenvalues of the matrix $L_{ij}$ determine the principal Lyapunov exponent  
 \begin{eqnarray}
  \lambda= \lim_{t \xrightarrow{} \infty} \frac{1}{t} log\left(\frac{L_{ij}(t)}{L_{ij}(0)} \right).
 \end{eqnarray}
Then the eigenvalues of the Jacobi matrix give the Lyapunov exponent. The positive Lyapunov exponent indicates the presence of chaos in the system \cite{cardosoGeodesicStabilityLyapunov2009a,
robinson1998dynamical}.
\color{black}
\subsection{Jacobi matrix method}
In this section, we will employ the Jacobi matrix method to calculate the principal Lyapunov exponent \cite{Gao:2022ybw}.
%yuBoundLyapunovExponent2022,,Cornish:2003ig
We consider a vortex (test particle) of unit mass in the equatorial plane $(\theta=\frac{\pi}{2})$ of the slow-rotating black hole whose Lagrangian is given by
% To find the Lyapunov exponent we shall apply the Jacobi method of eigenvalues following the paper \cite{Gao:2022ybw}. The Lagrangian is given by 
\begin{align}
   2\mathcal{L} =\sqrt{3} c_{s}^2  \Bigg[ - \mathcal{F}(r) \dot t^2 + \frac{1}{\mathcal{F}(r)} \dot r^2 + r^2 \dot \phi^2 - \frac{4 M a}{r} \dot t \dot \phi  \Bigg],   
\end{align}
where $\dot x= \Big( \frac{dt}{d\tau}, \frac{dr}{d\tau}, \frac{d \phi}{d \tau} \Big)$. 
On setting, 
%Let's set 
$\sqrt{3} c_{s}^2 =1$.
%for the rest of the paper.\\
then the generalised momenta are given by 
\begin{eqnarray}
    -p_{t} &&=   \mathcal{F}(r) \frac{dt}{d \tau} + \left(\frac{2 a M}{r}\right) \frac{d \phi}{d \tau}, \\
    p_{r} &&= \frac{\dot r}{\mathcal{F} (r)} ,\\
    p_{\phi} &&= -\left(\frac{2 a M}{r}\right) \frac{d t}{d \tau} + r^{2} \frac{d \phi}{d \tau},
\end{eqnarray}
Therefore, the Hamiltonian of the particle-like vortex is 
\begin{align}
  \mathcal{H}=-\frac{2 a M p_t p_{\phi }}{r^3 \mathcal{F}(r)}-\frac{p_t^2}{2 \mathcal{F}(r)}+\frac{1}{2} \mathcal{F}(r) p_r(\tau ){}^2+\frac{p_{\phi }^2}{2 r^2},
\end{align}
Using the Hamiltonian equation of motions, we obtain 
\begin{align}
  \dot t &=-\frac{2 a M p_{\phi }+r^3 p_t}{r^3 \mathcal{F}(r)},\\
 \dot p_{t} &= 0 ,\\
  \dot r &= p_{r} \mathcal{F}(r), \\
 \dot p_{r}&=\frac{1}{2 r^4 \mathcal{F}(r)^2} \Bigg(2 \mathcal{F}(r) p_{\phi } \left(r \mathcal{F}(r) p_{\phi }-6 a M p_t\right)- \nonumber \\&
 r \mathcal{F}'(r) \left(4 a M p_t p_{\phi }+r^3 \left(\mathcal{F}(r)^2 p_r^2+p_t^2\right)\right) \Bigg), \\
 \dot \phi&=  \frac{r p_{\phi }-\frac{2 a M p_t}{\mathcal{F}(r)}}{r^3},\\
  \dot p_{\phi} &= 0,
    % \dot t &= \frac{3 r^3 p_{t}-2 a M p_{\phi}}{r^3 \mathcal{F}(r)} \\
    % \dot p_{t} &= 0 \\
    % \dot r &= p_{r} \mathcal{F}(r) \\
    % \dot p_{r}&= \frac{1}{2} \left(\frac{ p_{t} \mathcal{F}'(r) \left(3 p_{t}-\frac{4 a M p_{\phi}}{r^3}\right)}{\mathcal{F}(r)^2}-
    % p_{r}^2 \mathcal{F}'(r) +\frac{2 p_{\phi}^2}{r^3} \right) \nonumber \\&
    % -\frac{6 a M p_t p_{\phi }}{r^4 \mathcal{F}(r)} \\
    % \dot \phi&= \frac{p_{\phi}}{r^2}-\frac{2 a M p_{t}}{r^3 F(r)} \\
    % \dot p_{\phi} &= 0
\end{align}
where $\prime$ denotes the derivative w.r.t. coordinate $r$.
We further define
\newline
%\newpage
% \begin{widetext}
% \begin{align}
 % F_{1} \equiv \frac{\dot r}{\dot t}\nonumber\\
 % =&-\frac{r^3 \mathcal{F}(r)^2 p_r}{2 a M p_{\phi }+r^3 p_t}  \label{F1}, \\
 % F_{2} \equiv \frac{\dot p_{r}}{\dot t} \nonumber \\
 % =&-\frac{1}{2 r \mathcal{F}(r) \left(2 a M p_{\phi }+r^3 p_t\right)}  \Big(2 \mathcal{F}(r) p_{\phi } \left(r \mathcal{F}(r) p_{\phi } \nonumber\\ &\quad
 % -6 a M p_t\right)-r \mathcal{F}'(r) \left(4 a M p_t p_{\phi }+r^3 \left(\mathcal{F}(r)^2 p_r^2+p_t^2\right)\right) \Big),\nonumber \\&& 
 % \label{F2}
  % F_{1} \equiv \frac{\dot r}{\dot t}&=&\frac{r^3 p_{r} \mathcal{F}(r)^2}{3 r^3 p_{t} -2 a M p_{\phi}} \label{F1} \\
  % F_{2} \equiv \frac{\dot p_{r}}{\dot t} &=& \frac{r \mathcal{F}'(r) \left(r^3 \left(3 p_t^2-\mathcal{F}(r)^2 p_r^2\right)-4 a M p_t p_{\phi }\right)+2 \mathcal{F}(r) p_{\phi } \left(r \mathcal{F}(r) p_{\phi }-6 a M p_t\right)}{2 r \mathcal{F}(r) \left(3 r^3 p_t-2 a M p_{\phi }\right)} \nonumber \\&& 
 % F_{2} \equiv \frac{\dot p_{r}}{\dot t} &=&\frac{r^3 \mathcal{F}(r) %\left(\frac{p_{t} \mathcal{F}'(r) \left(3 p_{t}-\frac{4 a M p_{\phi}}
 %{r^3}\right)}{\mathcal{F}(r)^2}- p_{r}^2 \mathcal{F}'(r)+\frac{2 %p_{\phi}^2}{r^3}\right)}{2 \left(3 r^3 p_{t}-2 a M p_{\phi}\right)}
  %\label{F2}
% \end{align}

\begin{align}
  F_{1} \equiv \frac{\dot{r}}{\dot{t}} 
  &= -\frac{r^3 \mathcal{F}(r)^2 p_r}{2 a M p_{\phi} + r^3 p_t} \label{F1}, \\
  F_{2} \equiv \frac{\dot{p}_{r}}{\dot{t}} 
  &= -\frac{1}{2 r \mathcal{F}(r) \left(2 a M p_{\phi} + r^3 p_t\right)} \Biggl( 2 \mathcal{F}(r) p_{\phi} \Bigl(r \mathcal{F}(r) p_{\phi} \nonumber \\
  &\quad - 6 a M p_t \Bigr) - r \mathcal{F}'(r) \biggl(4 a M p_t p_{\phi} \nonumber \\
  &\quad + r^3 \Bigl(\mathcal{F}(r)^2 p_r^2 + p_t^2\Bigr)\biggr) \Biggr), \label{F2}
\end{align}
% \end{widetext}
The normalization condition of the ``three-velocity" of the massive particle $g_{\mu \nu} \dot x^{\mu} \dot x^\nu=-1$ provides the following constraint 
\begin{eqnarray}
    r^4 \mathcal{F}(r) \left(r^2 \mathcal{F}(r) p_r^2+p_{\phi }^2+r^2\right)=r^3 p_t \left(4 a M p_{\phi }+r^3 p_t\right),%
    \nonumber \\
   % 4 a M p_{\phi} \left(a M p_{\phi}+ p_{t} r^3\right)+r^4 \mathcal{F}(r) \left( p_{r}^2 r^2 \mathcal{F}(r)+ p_{\phi}^2+r^2\right)= p_{t}^2 r^6 &
\end{eqnarray}
%we should write the below line in different way
then employing this constraint in the eq (\ref{F1}) and eq (\ref{F2}) to eliminate the $p_{t}$ gives 
% \begin{widetext}
\begin{align}
  F_{1} &= -\frac{r^3 \mathcal{F}(r)^2 p_r}{\sqrt{r^4 \mathcal{F}(r) \left(r^2 \mathcal{F}(r) p_r^2 + p_{\phi}^2 + r^2\right)}}, \label{eq:F1} \\
  F_{2} &= \frac{1}{2} \Biggl( \frac{12 a M p_{\phi}}{r^4} + \frac{p_{\phi}^2 \left(r \mathcal{F}'(r) - 2 \mathcal{F}(r)\right)}{\sqrt{r^4 \mathcal{F}(r) \left(r^2 \mathcal{F}(r) p_r^2 + p_{\phi}^2 + r^2\right)}} \nonumber \\
  &\quad + \frac{r^3 \left(2 \mathcal{F}(r) p_r^2 + 1\right) \mathcal{F}'(r)}{\sqrt{r^4 \mathcal{F}(r) \left(r^2 \mathcal{F}(r) p_r^2 + p_{\phi}^2 + r^2\right)}} \Biggr), \label{eq:F2}
\end{align}
Using the constraint $p_{r}=\cfrac{d p_{r}}{d t}=0$ for the equilibrium orbit of the particle yields
%Finally, we find the Lyapunov exponent using the Jacobian matrix $K_{ij}$ in the phase space $(r,p_{r})$ for the motion of the particle in the equilibrium orbit with $p_{r}=0$
% \begin{widetext}
\begin{align}
  K_{11} &= 0 , \\
  K_{22} &= 0, \\
  K_{12} &= -\frac{\mathcal{F}(r) \sqrt{r^4 \mathcal{F}(r) \left(p_{\phi}^2 + r^2\right)}}{r \left(p_{\phi}^2 + r^2\right)}, \\
  K_{21} &= \frac{1}{4 r \left(r^4 \mathcal{F}(r) \left(p_{\phi}^2 + r^2\right)\right)^{3/2}} \Biggl[ 
    r^6 \mathcal{F}'(r)^2 \left(-\left(p_{\phi}^2 + r^2\right)^2\right) \nonumber \\
  &\quad + 2 r^2 \mathcal{F}(r)^2 \left(6 r^4 p_{\phi}^2 + 4 r^2 p_{\phi}^4\right) \nonumber \\
  &\quad + 2 \mathcal{F}(r) \left(p_{\phi}^2 + r^2\right) \biggl(-48 a M p_{\phi} \sqrt{r^4 \mathcal{F}(r) \left(p_{\phi}^2 + r^2\right)} \nonumber \\ % Fixed: Added closing `}` for \sqrt
  &\quad + r^5 p_{\phi}^2 \Bigl(r \mathcal{F}''(r) - 2 \mathcal{F}'(r)\Bigr) + r^8 \mathcal{F}''(r)\biggr) \Biggr], \label{eq:K21}
\end{align}

and then finally evaluating the eigenvalues of the matrix
\begin{align}
\begin{pmatrix}
0 & K_{12}\\
K_{21} & 0 
\end{pmatrix},
\end{align}
gives the following (squared of) Lyapunov exponent for the timelike circular motion
\begin{align}\label{lambda2}
\lambda^2 &= -\frac{1}{2 r^6 \left(l^2 + r^2\right)}\Biggl(\mathcal{F}(r) \Big(-48 a l M \sqrt{r^4 \mathcal{F}(r) \left(l^2 + r^2\right)} \nonumber\\
&\quad + r^6 \left(l^2 + r^2\right) \mathcal{F}''(r) - 2 l^2 r^5 \mathcal{F}'(r)\Big) \Biggr)\nonumber \\
&\quad + \frac{1}{4} \mathcal{F}'(r)^2 - \frac{\mathcal{F}(r)^2 \left(2 l^4 + 3 l^2 r^2\right)}{r^2 \left(l^2 + r^2\right)^2},
\end{align}
where we have set the $p_\phi \equiv l$, the angular momentum along $\phi$ direction. Notice that the presence of parameters $a, ~l$ and $\xi$ is significant since they make a nontrivial contribution to the Lyapunov exponent. 
%\color{blue}
%We have noticed that, at the horizon, the universality of the chaos bound does not hold in our case. This issue is also examined in detail in \cite{Giataganas:2021ghs}.
%\color{black}
The stability of the equilibrium circular orbits can be determined by $\lambda^2$. It can be established that for the circular motion to be unstable, it should correspond to $\lambda^2 > 0$ \cite{leiCircularMotionCharged2022}, which thus indicates the presence of chaos. Our goal is to examine and if possible to establish the chaos bound $\lambda < \kappa$. On rewriting the bound eq (\ref{bound}) as 
\color{black}
\begin{eqnarray}
    \lambda^2 - \kappa^2 \le 0.
\end{eqnarray}
Thus the sign of $\kappa^2 - \lambda^2 $ would decide whether the bound is satisfied. Due to the complex expression of the $\lambda$, we resort to the numerical calculation and checking of the bound presented in the next section. \\

\subsection{Numerical Analysis}
%The chaos bound in near-horizon is analysed through the expansion of the Lyapunov exponent at the horizon \cite{}.
Here, we study how different parameters affect the chaos bound, identifying the spatial regions where the bound is respected or violated if any. The region we are concerned about is not limited to the near-horizon region, but also at a certain distance from the horizon i.e. in the vicinity of the horizon.

To analyse the bound $\kappa^2 - \lambda^2 $, we first find the equilibrium orbit $r_{0}$ using $p_{r}=F_{2}=0$ numerically at certain fixed parameter values, and then obtain the corresponding $\kappa^2-\lambda^2$ value by numerical calculation using eq (\ref{lambda2}), eq (\ref{eqF}) and eq (\ref{kappa}). To this end, without loss of generality, we set 
%Note that while using the formula for (squared of) Lyapunov exponent, the quadratic terms in the parameters from eq. 44 should also be taken into account. 
\begin{eqnarray}
    M=0.1. ~~
\end{eqnarray}
The presence of unstable equilibrium points would depend on the parameter values of $\xi, ~~ a,$ and $ ~l$. In Table \ref{tab:table1}-\ref{tab:table4} we have shown the position $r_{0}$ of the circular orbits at different parameter values. Then in the following analysis, we examine how the angular momentum, rotation parameter and $\xi$ are influencing the exponent and the chaos bound for extreme LTABH and the non-extrermal cases.
%Thus we shall analyse the behaviour of the chaos bound for various parameters: $\xi, ~a, ~l$. 
%The plots of the effective potential $V_{eff}$ at certain fixed parameters are shown in figure \ref{}.
%Figure \ref{lambda-kappa} is one such plot. 
\subsubsection{Extremal cases}
\begin{figure}[b]
        \centering
        \includegraphics[width=\linewidth]{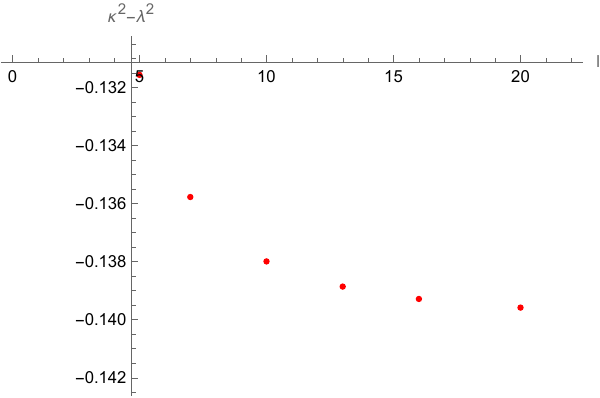}
        \caption{The plot for $\kappa^2 - \lambda^2$ for extremal LTABH at fixed parameters $a = M =0.1 ~~ \xi \xrightarrow{} 4 $ for different values of $l= 5,7, 10,13, 16, 20$.}
        \label{fig:lambda-kappa-a-extremal}
\end{figure}

Here, since at extremality, the temperature of the black hole is zero therefore $\kappa=0$ and the bound 
\begin{align}
    \kappa^2 - \lambda^2 \ge 0, 
\end{align}
reduces to
\begin{align}
    -\lambda^2 \ge 0 .
\end{align}
However, we observe that
%when there exists an unstable equilibrium $r_{0}$ (local maximum),
$\lambda^2 >0$ for the equilibrium point (more accurately the unstable equilibrium point) in the vicinity of the horizon and therefore the bound is violated (see Fig \ref{fig:lambda-kappa-a-extremal}). Moreover, from the table \ref{tab:table1}, it is observed that as we increase the angular momentum the position of the orbits slowly decreases and the motion becomes more chaotic.
\newline
\begin{table*}[th!]
\centering
\caption{\label{tab:table1}
Positions of the circular orbits of the particle around extreme LTABH at different values of $l$ when $M=a=0.1$ and $\xi \xrightarrow{}4$. The event horizon is located at $r_{1}= 0.4$. }
% \begin{ruledtabular} % <-- Remove this
\begin{tabular}{cccccccc}
\toprule % <-- Use booktabs top rule
$l$ &\multicolumn{1}{c}{5}&
\multicolumn{1}{c}{7} &
\multicolumn{1}{c}{10} &
\multicolumn{1}{c}{13} &
\multicolumn{1}{c}{16} &
\multicolumn{1}{c}{20} &
%\textrm{l}
\\
\midrule % <-- Use booktabs mid rule (replaces \hline)
$r_{0}$ & 0.962649 & 0.953677 & 0.949045 & 0.947253 & 0.946375 & 0.945763
%10 & 20 & 30 & 40\\
%100 & 200 & 300.0 & 400\\
\\ % Added a \\ here for spacing before bottomrule if needed, often not required.
\bottomrule % <-- Use booktabs bottom rule
\end{tabular}
\end{table*}
\begin{comment}
\begin{table}[t]
\caption{\label{tab:table1}
Positions of the circular orbits of the particle around extreme LTABH at different values of $l$ when $M=a=0.1$ and $\xi \xrightarrow{}4$. The event horizon is located at $r_{1}= 0.4$. }
\begin{ruledtabular}
\begin{tabular}{cccccccc}
%\textrm{Left\footnote{Note a.}}&
%\textrm{Centered\footnote{Note b.}}&
$l$ &\multicolumn{1}{c}{5}&
\multicolumn{1}{c}{7} &
\multicolumn{1}{c}{10} &
\multicolumn{1}{c}{13} &
\multicolumn{1}{c}{16} &
\multicolumn{1}{c}{20} &
%\textrm{l}
\\
\colrule
$r_{0}$ & 0.962649 & 0.953677 & 0.949045 & 0.947253 & 0.946375 & 0.945763
%10 & 20 & 30 & 40\\
%100 & 200 & 300.0 & 400\\
\end{tabular}
\end{ruledtabular}
\end{table}
\end{comment}
%depending on the parameter values. 
% In Fig \ref{fig:lambda-kappa-a-extremal}, with increase in $l$ the $\kappa^2 - \lambda^2$ becomes more and more negative reflecting the stronger violation of the bound.

% \newline
\subsubsection{Non-extremal cases}
\begin{figure}[b]
    \begin{minipage}{0.45\textwidth}
        \centering
        \includegraphics[width=\linewidth]{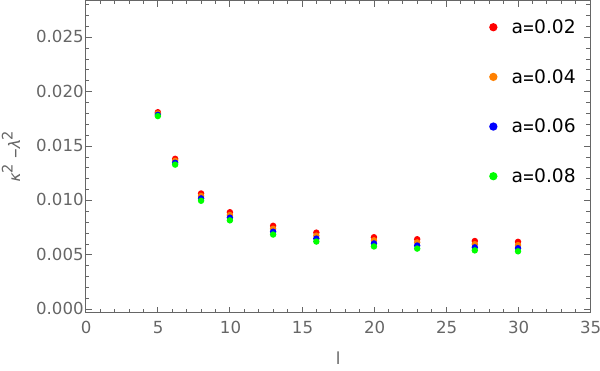}
         \caption{The figure shows the plot of $\kappa^2 - \lambda^2$ for different values of $l$ for $M=0.1$ at $~\xi=6$ and fixed $a$. }
%         %\caption{For $M=0.1,~~ L=5,10,20,30,$ at fixed $a=0.02,0.04,0.06,0.08$ (red,orange,blue,green) , $~\xi=6$}
         \label{fig:lambda-kappa-l-fixed-at-small-a}
    \end{minipage}
    \end{figure}

For non-extremal scenarios of the acoustic black hole, the various observed behaviour are summed up as follows 
% From Figures \ref{lambda-kappa-2}-\ref{lambda-kappa-4} various observed behaviour are summed up as follows 
\begin{itemize}
    \item When the rotation parameter $a$ and parameter $\xi$ are kept fixed, different values of angular momentum $l$ give rise to different values of the Lyapunov exponent, see Fig \ref{fig:lambda-kappa-l-fixed-at-small-a}. Further, the larger the value of $l$, the greater the value of $\lambda$ which implies stronger chaos in the system. However, the value of $\lambda$ does not exceed the bound.
    %The bound is violated at small a and small l. However, the bound is restored on increasing a and $l$ sufficiently.
    \item With fixed (say) $l=12$, different values of the parameter $\xi$ produce the different Lyapunov exponent corresponding to fixed rotation parameter $a$. Moreover, for larger as well as smaller  $\xi$ at fixed $a$, there is still no violation of the bound. See Fig. \ref{fig3:lambda-kappa-xi-fixed-l,a} (top panel). From the figure, we also see that the values of the $\lambda$ are close to each other even for varying $a$, and therefore $\kappa^2 - \lambda^2$ values are quite overlapping.
    \item When $\xi$ remains fixed, quantity $\kappa^2 - \lambda^2$ decreases as $a$ increases. On the other hand, when $l$ is fixed, $\kappa^2 - \lambda^2$ increases with an increase in $a$.
    % This feature matches with the Schwarzchild metric. [?]
    % we see no bound violation, however, the bound gets violated at small a and small $\xi$ values. 
     \item Finally, in Fig \ref{fig3:lambda-kappa-xi-fixed-l,a} (bottom panel) with fixed rotation parameter, different angular momentum corresponds to different Lyapunov exponent values. Even if $l$ increases sufficiently, we observe no bound violation. 
     \item Again similar to the extremal case, for the equilibrium point in vicinity of the horizon we observe the positivity of the $\lambda^2$. However, the bound is not violated.
     %the bound is violated for small values of $\xi$ and l. Again, we obtain different values of the Lyapunov exponent. Fig \ref{fig-overall:lambda-kappa-l-fixed-at-small-a}.
     
    %\item For the small values of a and l, the Lyapunov exponent becomes large enough that we observe to our surprise, the bound violation. \ref{}
\end{itemize}
 In non-extremal cases, numerically we see $\lambda^2 > 0$ for equilibrium points in the vicinity of the horizon, however far away from the horizon we observe $\lambda^2 <0$. Further there exist points $r_{0}^{\ast}$ ($ < r_{1}$) such that the positivity of $\lambda^2$ is seen to be in region $(r_{1}$, $r_{0}^{\ast})$.
%  Further there exist points $r_{0}^{min}$ ($ \le r_{1}$) and $r_{0}^{max}$ ($ \ge r_{1}$) such that the positivity of $\lambda^2$ is seen to be in region $(r_{0}^{min}$, $r_{0}^{max})$. 
% A schematic diagram is shown depicting this behaviour in Figure/Table \ref{}.
Also, we observe that as we move very far away (i.e. asymptotic limit) from point $r_{0}^{\ast}$, the $\lambda^2$ approaches zero.
Further, the bound is satisfied at all unstable equilibrium points.
%We then examine the impact of mass M on the chaos bound. The results with the fixed ... are shown in Fig \ref{}.\\
Note that similar behaviour was also observed for counter-rotating waves (negative $l$).\\ 
\\
Thus, we observe no bound violation throughout the scan of the parameter space. %This suggests that the slow-rotating acoustic black hole behaves similar to a real black hole, showcasing chaos while also validating the bound $\lambda \le 2 \pi T$ depending on the parameters. 
%This illustrates that the slow-rotating acoustic black hole acts similar to the real black hole where not only the chaos is present but also satisfies the bound $\lambda < 2 \pi T$.
Similar observations related to the non-rotating acoustic black hole have also been reported earlier in the literature \cite{wangGeometryOutsideAcoustic2020} where they have observed the saturation of the bound i.e. $\lambda= 2 \pi T_{H}$. However, in the slow-rotating ABH the chaos bound is not saturated even at the horizon $r_{1}$ which can be seen from using eq (\ref{lambda2}) and eq (\ref{kappa}).\\
% \color{red} 1. \st{ More points are needed to add in the plot 6 and plot 7(top and bottom panel) so that it looks continuous. Parameters($\xi, ~ a,~l$) etc can be chosen so that plot points distinguish from each other, if possible. }\\
% 2. Futher analytics regarding the bound in the rotating cases need to be investigated. \\
% 3. Look for more paper where the violation of the bound is observed/reported in rotating cases.\\
% 4. \st{People who have mailed us; those citations needs to added with proper modifications in the intro.} 
% \color{black}
%\newpage
\begin{table*}[t!]
\centering
\caption{\label{tab:table2} Positions of the circular orbits of the particle around non-extreme LTABH at different values of $l$ and $a$ at $\xi=6$. The event horizon is located at $r_{1}= 0.94641$.}
\begin{tabular}{cccccc}
\toprule % <-- Use booktabs top rule
 &\multicolumn{1}{c}{l}&\multicolumn{1}{c}{5}&\multicolumn{1}{c}{10} &\multicolumn{1}{c}{20}&\multicolumn{1}{c}{30}
% Ion&1st alternative&2nd alternative&lst alternative
%&2nd alternative
\\ 
\midrule % <-- Use booktabs mid rule (replaces \hline)
           & a=0.02 & 1.59189 & 1.54205 & 1.53060 & 1.52851 \\
    $r_{0}$ & a=0.04 & 1.59894 &1.54851&  1.53692 & 1.53481 \\
           & a=0.06 &  1.60598 &1.55494&   1.54322 & 1.54108\\
           & a= 0.08 & 1.61298 & 1.56135& 1.54949  & 1.54733 \\
\bottomrule % <-- Use booktabs bottom rule
\end{tabular}
\end{table*}
\newline
\begin{table*}[t!]
\centering
\caption{\label{tab:table3} Positions of the circular orbits of the particle around non-extreme LTABH at different values of $\xi$ and $a$. The corresponding event horizon is located at $r_{1}$ for $l=12$.}
\begin{tabular}{ccccc}
\toprule
         &       & \multicolumn{3}{c}{$\xi$}          \\
\cmidrule(lr){3-5}
 & $a$  & 6        & 10       & 12        \\ 
\midrule
$r_{1}$  &      & 0.94641  & 1.77460  & 2.17980   \\
\midrule
          & 0.02 & 1.53734  & 2.80220  & 3.45145   \\
$r_{0}$  & 0.04 & 1.54374  & 2.80537  & 3.45405   \\
          & 0.08 & 1.55647  & 2.81169  & 3.45924   \\
\bottomrule
\end{tabular}
\end{table*}
 \newline
\begin{table*}[t!]
\centering
\caption{\label{tab:table4} Positions of the circular orbits of the particle around non-extreme LTABH at different values of $\xi$ and $l$. The corresponding event horizon is located at $r_{1}$ for $a=0.02$.}
\begin{tabular}{ccccc}
\toprule
         &       & \multicolumn{3}{c}{$\xi$}          \\
\cmidrule(lr){3-5}
 & $l$  & 6        & 10       & 12        \\ 
\midrule
$r_{1}$  &      & 0.94641  & 1.77460  & 2.17980   \\
\midrule
          & 5    & 1.59188  & 3.15920  & 4.24152   \\
$r_{0}$  & 10   & 1.54205  & 2.82836  & 3.50026   \\
          & 20   & 1.53059  & 2.76589  & 3.38527   \\
\bottomrule
\end{tabular}
\end{table*}

\begin{figure}[t!]
     % \begin{minipage}{0.45\textwidth}
     %     \centering
     %     \includegraphics[width=\linewidth]{corrected-lambda-kappa-lat-fixed-a-small.pdf}
     %     \caption{For $M=0.1,~~ L=5,10,20,30,$ at fixed $a=0.02,0.04,0.06,0.08$ (red,orange,blue,green) , $~\xi=6$}
     %     \label{fig:lambda-kappa-l-fixed-at-small-a}
     % \end{minipage}\hfill
  \begin{minipage}{0.45\textwidth}
        \centering
        \includegraphics[width=\linewidth]{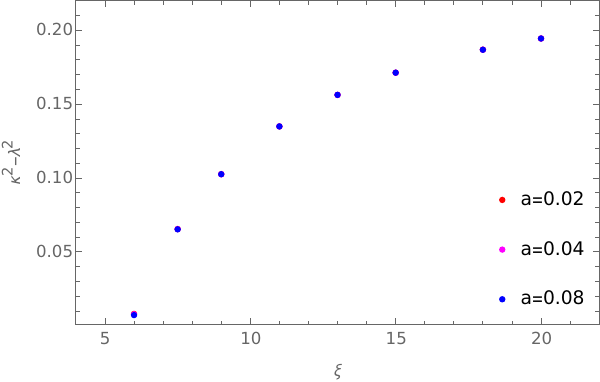}
        %\caption{For $M=0.1,~~ L=12, ~~ \xi=4,6,20$ with fixed $a=0.02,0.04,0.08$  (red, magenta, blue)}
        %\label{fig2:lambda-kappa-xi-fixed-a-l}
    \end{minipage}\par
    \vspace{2mm}
    % \caption{The left panel shows the plot of $\kappa^2 - \lambda^2$ for $M=0.1,~~ L=5,10,20,30,$ at fixed $a=0.02,0.04,0.06,0.08$ (red,orange,blue,green) , $~\xi=6$. The right panel shows the for $M=0.1,~~ L=12, ~~ \xi=6,9,15,20$ with fixed $a=0.02,0.04,0.08$  (red, magenta, blue).}
    \label{fig-overall:lambda-kappa-l-fixed-at-small-a}
    
     \begin{minipage}{0.45\textwidth}
        \centering
        \includegraphics[width=\linewidth]{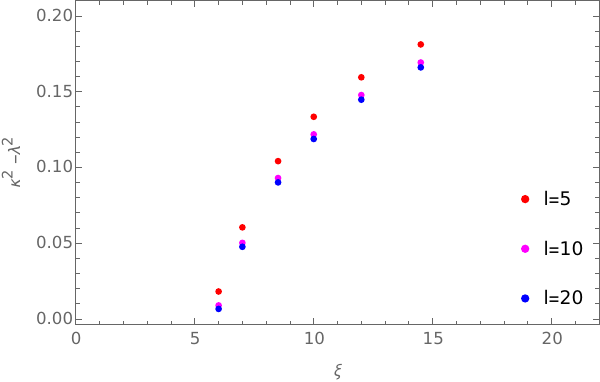}
         \caption{%The top panel shows the plot of $\kappa^2 - \lambda^2$ for different values of $l$ for $M=0.1$ at $~\xi=6$ and fixed $a$. 
     The top panel shows the plot of $\kappa^2 - \lambda^2$ for $M=0.1,~~ l=12$ for different values of $\xi$ with fixed $a$. The bottom panel shows plot of $\kappa^2 - \lambda^2$ for  different values of $\xi$ and  $M=0.1$,~~$a=0.02$ at fixed $l$. }
       % \caption{The top panel shows the for $M=0.1,~~ l=12, ~~ \xi=6,9,15,20$ with fixed $a=0.02,0.04,0.08$  (red, magenta, blue). The bottom panel shows the plot of $\kappa^2 - \lambda^2$ for different values of $\xi$ for $M=0.1$ and $a=0.02$, at fixed $~l$.}
        % \caption{For $M=0.1,~~ a=0.02~~ \xi=6,10,14.5$ at fixed $l=5,10,20$(red, magenta, blue)}
        \label{fig3:lambda-kappa-xi-fixed-l,a}
    \end{minipage}
\end{figure}

% \begin{figure}
%  \begin{minipage}{0.45\textwidth}
%     \centering
%     \includegraphics{at-small-a-0.02-lambda-kappa-xi-fixed-l.pdf}
%      \end{minipage}
%     \caption{Caption}
%     \label{fig:enter-label}
% \end{figure}
%below is the old:
% \begin{figure*}
%     \begin{minipage}{0.45\textwidth}
%         \centering
%         \includegraphics[width=\linewidth]{corrected-at-small-a-0.02-lambda-kappa-xi-fixed-l.pdf}
%         \caption{For $M=0.1,~~ a=0.02~~ \xi=6,10,14.5$ at fixed $l=5,10,20$(red, magenta, blue)}
%         \label{fig3:lambda-kappa-xi-fixed-l,a}
%     \end{minipage}\hfill
    
    % \begin{minipage}{0.45\textwidth}
    %     \centering
    %     \includegraphics[width=\linewidth]{lambda-kappa-l-fixed-a=0.1=M.pdf}
    %     \caption{}
    %     \label{fig:lambda-kappa-a}
    % \end{minipage}
    % \caption{something}
%\end{figure*}
\section{Conclusions}
In this study, we explored the dynamics of a vortex, treated as a unit-mass test particle, in the vicinity of a slow-rotating acoustic black hole (ABH) derived from the Gross-Pitaevskii and Yang-Mills theory. Our investigation centered on three key aspects: the effective potential governing vortex motion, the identification of circular orbits including the innermost stable circular orbit (ISCO), and the characterization of chaotic behavior via the Lyapunov exponent.

We derived the effective potential for the vortex and confirmed the existence of both stable and unstable circular orbits. The ISCO, analogous to that in Kerr black holes, was identified, with its radius influenced by the fluid’s rotational properties (parametrized by the rotation parameter \(a\)). This similarity underscores the power of acoustic black holes as analogues to gravitational systems, enabling possible laboratory exploration of relativistic phenomena.

To assess chaotic dynamics, we employed the Jacobi matrix method to compute the Lyapunov exponent (\(\lambda\)) for equilibrium orbits. In non-extremal ABHs (\(\xi > 4\)), where the surface gravity \(\kappa\) is non-zero, \(\lambda\) approaches \(\kappa\) for unstable equilibrium points near the acoustic horizon (\(r_1\)) but does not exceed it, satisfying the chaos bound \(\lambda \leq \kappa\). Numerical analysis across various parameter values (\(\xi\), \(a\), and angular momentum \(l\)) consistently showed no bound violation, reinforcing the stability of this limit in non-extremal cases. Conversely, in extremal ABHs (\(\xi = 4\)), where \(\kappa = 0\), the bound is violated as \(\lambda^2 > 0\), indicating enhanced chaotic behavior. This violation highlights a distinct dynamical regime in extremal configurations, where the absence of surface gravity permits unbounded exponential divergence of trajectories.

These findings advance our understanding of vortex dynamics in slow-rotating acoustic black holes and their parallels with gravitational black holes. The respect of the chaos bound in non-extremal cases aligns with previous studies of acoustic systems (e.g., \cite{wangGeometryOutsideAcoustic2020}), though unlike some non-rotating cases where \(\lambda\) saturates the Maldacena-Shenker-Stanford (MSS) bound (\(\lambda = 2\pi T_H\)), our slow-rotating ABH does not exhibit saturation at the horizon. The extremal case violation suggests that rotation and criticality introduce unique chaotic signatures, potentially linked to frame-dragging effects encoded in the Lense-Thirring metric.
 
In future, it would be interesting to develop a thermodynamic-like framework for slow-rotating ABHs, exploring quantities such as Gibbs free energy, specific heat, and statistical entropy, akin to studies of gravitational black holes \cite{Yasir:2023ysw}. This could elucidate phase transitions and thermal stability, enhancing the analogy with relativistic thermodynamics. Further, it would be interesting to investigate quantum corrections to the classical vortex dynamics and their impact on the chaos bound. Exploring the relationship between quantum chaos and analogue Hawking radiation could deepen connections to quantum gravity.
\section*{Acknowledgments}
 We would like to thank Prof. Kamal L. Panigrahi for valuable comments and suggestions during the preparations of the manuscript. 
 % We thank the anonymous referee for constructive suggestions which helped in modification of the paper.
%REV\TeX{}, offering suggestions and encouragement, testing new versions,
% \end{acknowledgments}
%\newpage
% \appendix*

% \nocite{*}

\appendix

% \section{Appendix section}\label{app}

\section{Constant $r$ geodesic}\label{APPENDIX}
%\begin{widetext}
% \color{red}
The slow-rotating acoustic metric eq (\ref{slow-rot}) can  be rewritten as
\begin{equation}
\begin{split}
    ds^2=-\mathcal{F}(r)dt^2+\frac{dr^2}{\mathcal{F}(r)}+r^2 \Big[d\theta^2+\sin^2{\theta} \big(d\phi-\frac{2 m a}{r^3}dt\big)^2 \Big], 
\end{split}
 \end{equation}
The four conserved quantities associated with this line-element are given by \cite{gray2022slowly}
\begin{subequations}
    \begin{equation}
        \mathcal{C}= \frac{1}{\sin^2 \theta} \left( \frac{d\phi}{d\tau}\right)^2 + \left( \frac{d\theta}{d\tau}\right)^2,\label{carter}
    \end{equation}
    \begin{equation}
        \epsilon= \frac{2 a l M}{r^3}+ \mathcal{F}(r) \frac{dt}{d\tau},
    \end{equation}
    \begin{equation}
         l=r^2 \sin^2{\theta} \left(\frac{d\phi}{d\tau}-\frac{2 a M}{r^3} \frac{dt}{d\tau}\right),
    \end{equation}
    
    \begin{equation}
        \begin{split}
           \delta= -\mathcal{F}(r) \left( \frac{dt}{d\tau}\right)^2 + \frac{1}{\mathcal{F}(r)}  \left( \frac{dr}{d\tau}\right)^2 +r^2 \mathcal{C} \\ -\frac{r^2}{\sin^2\theta}  \left( \frac{d\phi}{d\tau}\right)^2 +\frac{l^2}{r^2 \sin^2\theta}, 
        \end{split}
    \end{equation}
\end{subequations}
\\
where $\delta=-1$ for time like geodesic.
% %\color{cyan}
% On finding Y as follws
% (For expression of the $X(r)$ in the paper: arxiv-2202.09010 (eq number 2.18): find $\dot t$ from eq 2.9 and put it into the eq 2.10. )
% \begin{align}
%    -1=r^2 \left(-\csc ^2(\theta )\right) \left(\frac{2 a m \epsilon }{F r^3}+\frac{l \csc ^2(\theta )}{r^2}\right)^2-\frac{\left(r^3 \epsilon -2 a l m\right)^2}{F r^6}+c r^2+\frac{\left(r'\right)^2}{F}+\frac{l^2 \csc ^2(\theta )}{r^2}
% \end{align}
% On taking $\theta->\pi/2$ and putting the $\mathcal{F}$, we get
% \begin{align}
%     \dot r^2 &= Y(r) \nonumber \\
%     &=\frac{2 m \xi  \left(c r^2+1\right) (r-2 m)^2+r^2 \left(\left(c r^2+1\right) (2 m-r)+r \epsilon ^2\right)}{r^3}
% \end{align}
% Now, if we set the additional constraint: $ 1+ C r^2 >0 $, then we get $r> 2 m.$ \\
Now solving the above equations, we have the e.o.m for $r$ 
 \begin{equation}
        \begin{split}
  \Dot{r}^2=Y(r)
  =\frac{1}{r^3} \Bigg[ \epsilon \left(4 a l M \csc ^4(\theta )-4 a l M+\epsilon r^3\right)\\-r F(r) \left(\mathcal{C} r^4-l^2 \csc ^2(\theta )\left(\csc ^4(\theta )-1\right)+r^2\right)\Bigg].
   \label{req}
   \end{split}
\end{equation}
In this coordinate, the geodesic trajectory for $r$ depends on the angular coordinate $\theta$. Since we are interested in circular geodesics with constant radial coordinate $r=r_0$ we focus on the equatorial plane where $\theta=\frac{\pi}{2}$. 
Now eq (\ref{req}) reduces to ,
    \begin{equation}
        \begin{split}
           \Dot{r}^2 =\frac{1}{r^3}\Big[r^2 \left(\left(\mathcal{C} r^2+1\right)(2 M-r)+\epsilon^2 r\right)+2 M \xi \\\left(\mathcal{C} r^2+1\right) (r-2 M)^2 \Big]. 
        \end{split}
    \end{equation}
Now for constant $r$-orbits of time-like geodesics, we must have $Y(r_0)=0$ and $Y'(r_0)=0$ and the
stability of the geodesic depends on the sign of the $Y''(r_0)$. 
%we are interested in finding the location of possible constant-$r$ geodesics.
% \begin{widetext}
 \begin{align}
  Y(r_0) &= \frac{1}{r_0^3} \Biggl( r_0^2 \left(\left(\mathcal{C} r_0^2 + 1\right) \left(2M - r_0\right) + \epsilon^2 r_0\right) \nonumber \\
  &\quad + 2M\xi \left(\mathcal{C} r_0^2 + 1\right) \left(r_0 - 2M\right)^2 \Biggr), \label{a8} \\
  Y'(r_0) &= \frac{1}{r_0^4} \Biggl( -8M^3\xi \left(\mathcal{C} r_0^2 + 3\right) \nonumber \\
  &\quad + 2M(\xi + 1) r_0^2 \left(\mathcal{C} r_0^2 - 1\right) - 2\mathcal{C}r_0^5 + 16M^2\xi r_0 \Biggr), \\
  Y''(r_0) &= \frac{1}{r_0^5} \Biggl( 2 \left(8\mathcal{C}M^3\xi r_0^2 - \mathcal{C}r_0^5 + 48M^3\xi \nonumber \right. \\
  &\quad \left. - 24M^2\xi r_0 + 2M\xi r_0^2 + 2Mr_0^2\right) \Biggr).
\end{align}
% \end{widetext}
Now using the above equations, we have
\begin{equation}
\begin{split}
    \text{Sign}(Y''(r_0))=\text{Sign}\Big[\mathcal{C} r_0^4 (2 M (\xi +1)-3 r_0)+\\8 M^2 \xi 
   (3 M-r_0)\Big]. 
\end{split}
\end{equation}
From the above expression, it is clear that how the stability condition depends on 
$r_{0}$ M (mass), and $\xi$(a tuning parameter it determines the position of the inner and outer acoustic
event horizons). We can summarize the stability criteria \cite{Baines:2022srs} as: 
\begin{enumerate}
    \item   $\text{Sign}(Y''(r_0))=+$ if $r_0 < 3M$ and $r_0 <\frac{2 M (\xi +1)}{3}$,
    \item $\text{Sign}(Y''(r_0))=-$ if $r_0 > 3M$ and $r_0 >\frac{2 M (\xi +1)}{3}$,
\item If $r_0 $ lies between $\frac{2 M (\xi +1)}{3}$ and $3M$, 
the $\text{Sign}(Y''(r_0))$ will depend on their relative magnitudes.
\end{enumerate}
 Again solving eq (\ref{a8}) we get the expression for the Carter constant, and applying the positive condition on it we get 
 \begin{equation}   
 r_0 > 2M~~ and ~~\xi <\frac{r_0^2}{2 M r_0-4 M^2}.
 \end{equation}
 Hence we can conclude that there will be many possible constant $r$-orbits satisfying the above conditions.
 Again, the Carter constant in eq (\ref{carter}) can be reexpressed in the following form 
\begin{equation} 
   \mathcal{C}=\frac{4 a l M \epsilon}{r^5 \mathcal{F}(r)}+\frac{l^2}{r^4}=\cfrac{2  l \Omega \epsilon}{r^2 \mathcal{F}(r)}+\frac{l^2}{r^4},
   \end{equation} 
   where $\Omega$ is nothing but the dragging  parameter defined as   
   \begin{align}
       \Omega=\frac{g_{t\phi}}{g_{\phi\phi}}=\frac{2 M a}{r^3}.
\label{drag}
   \end{align}
and one can notice the presence of dragging parameter in the the Lyapunov exponent eq (\ref{lambda2}) as well.
This frame-dragging parameter $\Omega$ appears in the Lyapunov exponent, impacting the stability of orbits and reflecting how rotational 
effects influence the geodesic trajectories.
   \color{black}

\begin{comment}
\subsection*{Appendix subsection}

However, different customer classes may visit
stations in different orders; the system
is not necessarily ``feed-forward.''
We define the {\em path of class $k$ customers} in
as the sequence of servers
they encounter along their way through the network
and denote it by
\begin{equation}
\mathcal{P}=\bigl(j_{k,1},j_{k,2},\dots,j_{k,m(k)}\bigr). \label{path}
\end{equation}

Sample of cross-reference to the formula (\ref{path}) in\break \ref{app}.

\section{Appendix section}\label{appB}

We consider a sequence of queueing systems
indexed by $n$.  It is assumed that each system
is composed of $J$ stations, indexed by $1$
through $J$, and $K$ customer classes, indexed
by $1$ through $K$.  Each customer class
has a fixed route through the network of
stations.  Customers in class
$k$, $k=1,\ldots,K$, arrive to the
system according to a
renewal process, independently of the arrivals
of the other customer classes.  These customers
move through the network, never visiting a station
more than once, until they eventually exit
the system.    
\end{comment}

\end{document}